\documentstyle[manuscript,aps]{revtex}
\begin{document}
\draft
\title{Percolation in Cluster-Cluster Aggregation Processes}

\author{Anwar Hasmy$^{1,2}$ and R\'emi Jullien$^1$}

\address{$^1$Laboratoire de Science des Mat\'eriaux Vitreux, UA 1119 CNRS,
      Universit\'e Montpellier II, Place Eug\`ene Bataillon,
                     34095 Montpellier Cedex 5, France}

\address{$^2$Laboratoire de Physique des Solides,
Universit\'e Paris-Sud, Centre d'Orsay, 91405 Orsay, France}

\date{\today}
\maketitle

\begin{abstract}
Numerical simulations of Diffusion-Limited and
Reaction-Limited Cluster-Cluster
Aggregation processes of identical particles are
performed in a two-dimensional box.
It is shown that, for concentrations larger than a characteristic gel
concentration, the morphology of the resulting
spanning cluster at the gel time $t_g$
exhibits a crossover length $L_c$ between percolation
($l>L_c$) and aggregation ($l<L_c$). $L_c$ vanishes
when increasing $c$, and, at a critical concentration value $c_p$
(where $L_c \rightarrow 0$)
the entire spanning cluster scales as the percolating
cluster obtained by standard percolation. Even if for $c>c_p$
the long-range correlations are similar to that of percolation,
the vanishing links in the structure suggest that
an homogeneous regime appears at small scales.
\end{abstract}


\section{Introduction}

The aggregation processes of small particles have attracted a great theoretical interest
in the last decade \cite{fl,jb,me1,vi},
due to their wide range of applications.
On can distinguish Particle-Cluster aggregation,
whose prototype is the Witten-Sander
model \cite{ws}, and Cluster-Cluster aggregation \cite{me2,kbj}.
While the former model applies to field-induced growth processes, such as
electrodeposition \cite{ms,mm}, dielectric breakdown \cite{np},
viscous fingering \cite{ac,lf}, etc..., the latter explains
true aggregation processes such as polymerisation \cite{cdd,pbd},
oil in water emulsions \cite{bmg}, soot particles in flames \cite{cls},
flocculation of colloidal particles \cite{whl,bc,fpv,re,acw}.
Among Cluster-Cluster aggregation processes  one distinguishes diffusion-limited
cluster-cluster aggregation (DLCA) \cite{me2,kbj} and chemically-limited
(also called reaction-limited)
aggregation (RLCA) \cite{jk,kj,bb}
which correspond respectively to fast (fully screened) and slow (partially
screened) aggregation of colloids \cite{whl}.

The CCA processes leads to a flocculation regime when the
concentration is smaller than a characteristic
gel concentration $c_g$, and to a gelation regime otherwise. In
flocculation regime it remains a single fractal cluster
at the end of the aggregation process with a fractal dimension
$D$ equal to 1.45 in DLCA case (and 1.65 in RLCA case).
In the gelation
regime it appears an ``infinite'' cluster at a given time
$t_g$ (hereafter refered as the gelling cluster).
When working with a finite (but large)
box, such a cluster is usually defined
as touching the box from edge to edge as in the percolation theory \cite{st}.
Since both physical problems are
quite similar, it is tempting to explain the infinite cluster formation in DLCA
by the percolation theory which excludes all dynamical phenomena and this
has been done by several authors \cite{kh,hk,ko,jmh,nd,gi}.
The fractal dimension of the infinite cluster of DLCA has been
found to be different than the one of percolation \cite{kh,hk}
and it has been argued that this discrepancy is due to
dynamical effects. However, recent aggregation experiments \cite{whm}
on attractive coagulated
particles suggest that percolation transition occurs at the particle
concentration 0.42.
Therefore the question is still open: are CCA models
and percolation theory  compatible?

In this paper we show that the
infinite cluster obtained at $t_g$ in CCA models exhibits
a crossover length $L_c$ between  percolation and aggregation,
for concentration values $c$  larger than the gel concentration $c_g$.
Furthermore, we show that $L_c$ vanishes when increasing $c$, and
for $c \approx 0.5$  the percolation regime
shows up at all length scales in the infinite cluster. These results
suggest that CCA processes
can be viewed as an irreversible percolation
phenomenom, as the invasion percolation model without trapping \cite{ww,po} and
the model for a diffusion front\cite{sa}.

\section{The Model}

The two-dimensional Diffusion-limited Cluster-Cluster Aggregation model consists in a Monte Carlo
algorithm which builds clusters on a lattice within
a square box of edge length $L$. Initially
particles are distributed randomly
(but uniformily in the space) on the lattice sites up to volume fraction
(concentration) $c$.
In order to insure that the
diffusion coefficient of the clusters varies as the inverse
of their radius $R$, a particle (or a cluster) is chosen randomly
according to a probability:
\begin{eqnarray}
p_{n_i} = {n_i^{\alpha}\over \sum_i n_i^{\alpha}}
\label{E3}
\end{eqnarray}
where $\alpha$ (=$-1/D$) is the kinetical exponent.
Then the cluster performs a translational motion by one unit
(taking account Periodic Boundary Conditions, PBC) in any of the four
directions $\pm 1,\pm 1$ chosen at random. If the cluster does not
collide with another, the displacement is performed and the
algorithm goes on by choosing again another
cluster. If a collision occurs between two clusters they
stick together forming a new large cluster. In our simulation
we have considered that collision occurs when a particle of
a cluster tries to occupy a particle of another cluster. In that case
the cluster is not displaced but a bond is established
between the two contacting particles. If there are more than
one collision at a given motion, only
one bond is chosen at random. This
trick has used by Kolb in a reversible diffusion-limited cluster
aggregation model \cite{ko}. This important
variant implies that there are no loops (as in off-lattice DLCA \cite{he})
and that
there is no intrinsic percolation at the begining of the
process: the intitial concentration can be varied up to unity.

For concentrations larger than $c_g$ there exists a gel time $t_g$ where
a cluster  becomes infinite. This gelling cluster is stored for a
numerical analysis, and to compare, we leave the aggregation
process continuing up to the time where it remains only one cluster  (hereafter refered as
the final cluster).
In the case of Reaction Limited Cluster-cluster Aggregation \cite{jk,kj,bb},
in addition to the algorithm
described above, it is introduced a sticking probabilty $p (\ll 1)$.
after a collision a new bond is created only if a
random number (uniformly distributed
between zero and one) is smaller than $p$.
Here, we have performed two-dimensional simulations
in boxes of different sizes up to $L=$ 240.

\section{Results and Discussion}

Typical results that compare qualitatevely the
resulting morphologies of the gelling cluster (G) and
the final cluster (F) in DLCA are
are illustrated in figure 1. In the case of the gelling cluster
the other remaining
clusters have been discarded and therefore
are not shown. Note that the morphology of the cluster shown in figure 1a is
strongly reminiscent of that of a percolating
cluster as obtained in standard percolation theory \cite{st}
when occupying randomly the sites of a square
lattice with the percolation probability $p_c=0.59273$.
To make quantitative comparaisons between G and F,
we have calculated the mass $M$ dependent box size $l$
using the mass-counting algorithm \cite{fe}. In figure
2 we show the log-log plot of $M$ versus $l$ for
the same concentration than those considered in figure 1. The resulting
slope indicates that M scales as $l^{D_p}$ and $l^d$, for
G (open circles) and F (black squares),
respectively. $D_p$ (=1.89) is close to the fractal dimension
of the percolating cluster in two dimensions,
and $d$(=2) is the spatial dimension.

If the mass of the gelling cluster G or the mass of the final
cluster F scales as $l^D$
we might be tempted to conclude that the quantity $m={M(l)\over{c l^D}}$
should not depend on $l$. To test this we show,
in figure 3, the curves giving $m$ as a function of $l$.
Figures 3a and 3c corresponds to G for $D=D_p$
and D=d, respectively and figures 3b and 3d correspond to F for $D=D_p$
and D=d, respectively.
One observes on these figures that $m$ is independent
on $l$ only in cases 3a and 3d and for
sufficiently large concentrations, confirming our first
guess of different fractal dimensions, 1.89 and 2.00,
for G and F respectively.
Figures 3b and 3c are shown here as  conter-examples to
illustrate the high precisions on these estimates
for the fractal dimensions.
In general, i.e. for not too large concentrations, one
can define a $c$-dependent crossover length
$L_c$ such that $m$ becomes independent on $l$ only for
$l>L_c$. For $l<L_c$, another linear regime
is observed, which is better extended at very low concentration,
with a slope of about -0.4 and -0.5
in cases 3a and 3d, respectively, corresponding to the fractal dimension
$D\simeq 1.89-0.4\simeq 2.00-0.5\simeq 1.5$, close to the one of DLCA clusters.
Therefore $L_c$ defines a change of scaling regime between DLCA
for short lengths and either percolation
(G) or homogeneity (F) for large lengths.
In fact, $L_c$ should be proportional
to the characteristic length correlation $\xi$
(or average size) of DLCA fractal
aggregates \cite{he}. Note that in all cases $L_c$ vanishes for
$c \approx 0.5$, and this is the reason of the absence
of crossover  in the curves reported in figure 2.

When estimating the fractal dimension of the aggregates $D_{agg}$ in
the  range of lengths corresponding to the aggregation
regime ($l<L_c$), we found, in
both cases G and F, that it increases significantly
with concentration (the data
are reported in figure 4). Those results confirm previous
conclusions in two dimensions \cite{vpd} as well as in three dimensions
\cite{hj}. As a consequence
the $c$-dependence of $L_c$ cannot be annalysed as a simple scaling relation.
Moreover, the discrete values of $l$ that require mass-counting
calculations, and
finite size effects, impedes us to determine $L_c$ within a
sufficiently small range
of error.  Anyway, all the results depicted in figs. 2 and 3
suggest that, at least for
$l>L_c$ the mass of the gelling cluster G scales with $l$
as a standard percolation
cluster. However such results are not sufficient to insure
that their morphologies
are the same. It is well known that there exist additional
quantities (different and independent
on the fractal dimension) to characterize a percolating cluster.
In principle, the morphology is entirely characterized
by an infinite set of exponents \cite{st,is}, but
here
we shall focuse on two particular exponents: the fractal dimensions
of the ``backbone'' $D_{bb} \cite{hs}$
and the fractal dimension of the ``links'' \cite{ps}.
The backbone of a cluster is the ensemble which remains after removing
dead-ends (or dangling-ends). For a percolating cluster in two
dimension the fractal
dimension of the
backbone $D_{bb}$ is $1.61$. The links (also called red bonds) are
the sites of the cluster that are singly connected,
that is, if we take out a link the connection between
the entire cluster is broken.
For a  percolating cluster in two dimensions the fractal dimension
of links $D_l$ is equal to 0.75.

In order to identify the backbone and the links
of a given cluster, we have been obliged (due to the PBC
considered in the CCA simulations) to span the  cluster out of the
box, taking account of PBC. Then, we have applied
to the new cluster configuration
(with size larger than the box size $L$, in almost all situations)
a procedure suggested by Hermann et al. \cite{hhs} to identify the backbone and the links.
Finally, the resulting backbone (and links) is unspanned and
returned inside the original box.
In figures 1, the sites depicted with strong
grey and black colors denotes the backbone and the links,
repectively.

We have calculated the fractal dimension of the backbone
$D_{bb}$ for both  G and F clusters
and different
concentration values. The results are depicted in figure
5 for two box sizes length $L$ =90 (black symbols)
and 120 (open symbols).
Note that for small concentration
values the  $D_{bb}$ values, for both G and
F clusters, are quite similar. This can be understood, since,
as mentioned above, the fractal dimension in the aggregation regime
is almost the same
in both cases.
This conclusion is
more convincing for $L=120$, but for larger $c$ values,
$D_{bb}$ of the  gelling cluster G becomes smaller than
$D_{bb}$ of the final cluster F.
Furthermore, in the G case, $D_{bb}$ reaches the
value 1.61 (the same value than for a
percolating cluster) for $c \approx 0.5$. For $c>0.5$,
$D_{bb}$ saturates to 1.89 (the value of the fractal
dimension of a percolating cluster).
In the F case, the same kind of saturation phenomenon occurs but, for $c>0.5$,
$D_{bb}$ saturates to the spatial dimension $d$.
The increasing value of $D_{bb}$ suggests that the
backbone structure is mainly reflecting the
fractal aggregate  structure (Which also increases with concentration as shown
in figure 4).
It is only when the crossover length $L_c$ vanishes that
the backbone structure becomes characteristic of the
one of a percolating cluster, since  $D_{bb}$
varies from
1.61 to 1.89 as shown  in figure 5.

In figure 6 it is shown that for $c$ smaller than about 0.5, the
fractal dimension $D_l$ of the links is approximately equal to 1,
suggesting that the
mass of the links scales with $l$ in a trivial manner.
However, for
$c \approx 0.5$, $D_l$ reaches the value 0.75 as in a percolating cluster.
The $D_l$ undependence on the box size $L$ suggests
that there exist a threshold concentration, close to $0.5$
where the entire G cluster
scales exactly as a percolating cluster.
When $c$ is increased above this threshold, $D_l$ vanishes, similarly to
the percolation theory when the occupation probability
$p$   is increased above  $p_c$ \cite{is}.
The vanishing links in the gelling cluster for large $c$
values suggest that, at very small scales, the system
becomes homogeneous, even if, at large distances, the system
scales as a percolating cluster as shown  in figure 3.

In order to appreciate the degree of generality of our results,
we have also performed
some calculations in the reaction-limited case (RLCA).
In figure 7 we have reported the $m(l)$ curves and it can be shown that
they exhibit the same qualitative behavior than in the DLCA case (figure 3).
These results
suggest that the percolation scaling  could  exist
in other  kinds of CCA processes such as the ballistic-limited \cite{bj}, the
convection-limited \cite{wb} and the fluctuating bond \cite{jh} aggregation models.

\section{conclusion}

In this paper we have shown that the infinite cluster obtained at the gel time
$t_g$ in CCA models 
exhibits a crossover length $L_c$ between aggregation and percolation.
Moreover, $L_c$ vanishes at
a critical concentration $c_p (\simeq 0.5)$ where the mass of
the entire system (and its backbone and links) scales as for a percolating
cluster obtained at $p_c$. For $c>c_p$ the percolation regime persists
at least at large scales, because the vanishing links suggest that an homogeneous
regime apperas at small scales. The value obtained here for $c_p$ is close to
the value reported in the above mentioned experimental work \cite{whm}
($\simeq 0.42$), but typical statistical fluctuations to determine
a critical value impedes us to insure if there are some relation
between these two critical probabilities. It might also be worth finding a relation between
$c_p$ and the critical probability of the site percolation threshold ($p_c=0.59273$) on 
a square lattice. On the other hand, the fact that close to $c_p$ the fractal dimension of the links $D_l$
becomes equal to that of the percolating clusters suggests that the critical behavior in both
cases are the same, since
the correlation lenght exponent $\nu$ is given by $1 \over D_l$ \cite{co}.
Preliminary calculations on a cubic lattice suggest that the results reported here
are quite general and extend in three dimensions.

One of us (A. H.) would like to acknowledge support from CONICIT (Venezuela).
The numerical calculations where done on the computers of the CNUSC (Centre Universitaire
Sud de Calcul), Montpellier, France, with support from CNRS.

\begin{figure}
\caption{Typical configurations for a (a) gelling cluster G and (b) final cluster F,
for c=0.5 and L=120. The
sites shown in      dark grey and black
belong to the backbone and the links, respectively.}
\end{figure}

\begin{figure}
\caption{Log-Log plot of $M$ versus $l$ for $c$=0.5 and $L$=240, for the gelling
cluster G (open circles) and the final cluster F (black squares). These curves
result from averages over 40 simulations.}
\end{figure}

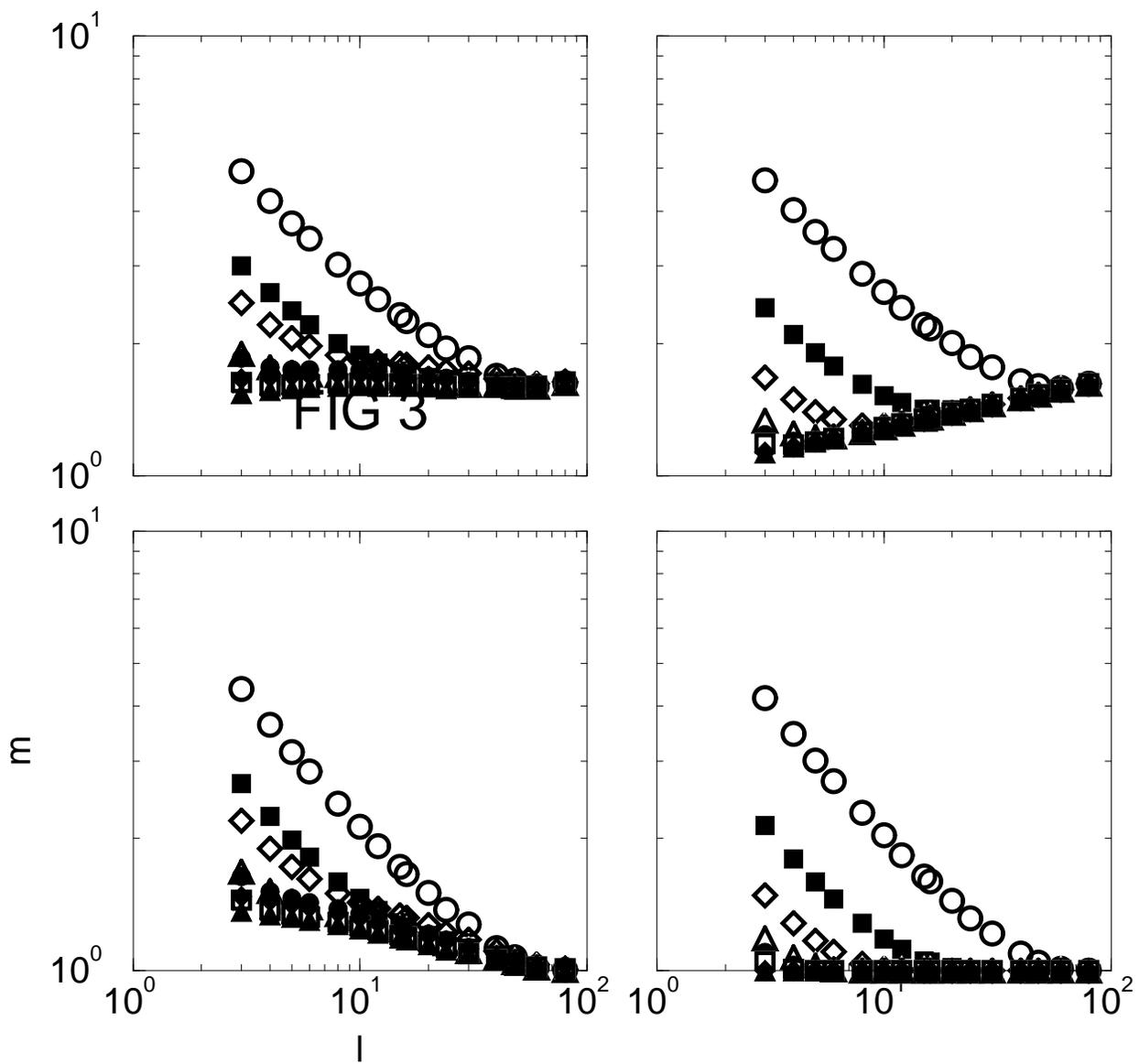
\begin{figure}
\caption{Log-Log plot of $m(={M(l) / l^D})$ versus $l$ for $L=240$ and
different concentration values (0.1, 0.2, 0.3,..., 0.8, from top to bottom)
in the DLCA case.
(a) and (c) correspond to G-cluster for $D=D_p$ and $D=d$, respectively.
(b) and (d) correspond to F-cluster for $D=D_p$ and $D=d$, respectively.
These curves result from averages over 40 simulations.}
\end{figure}

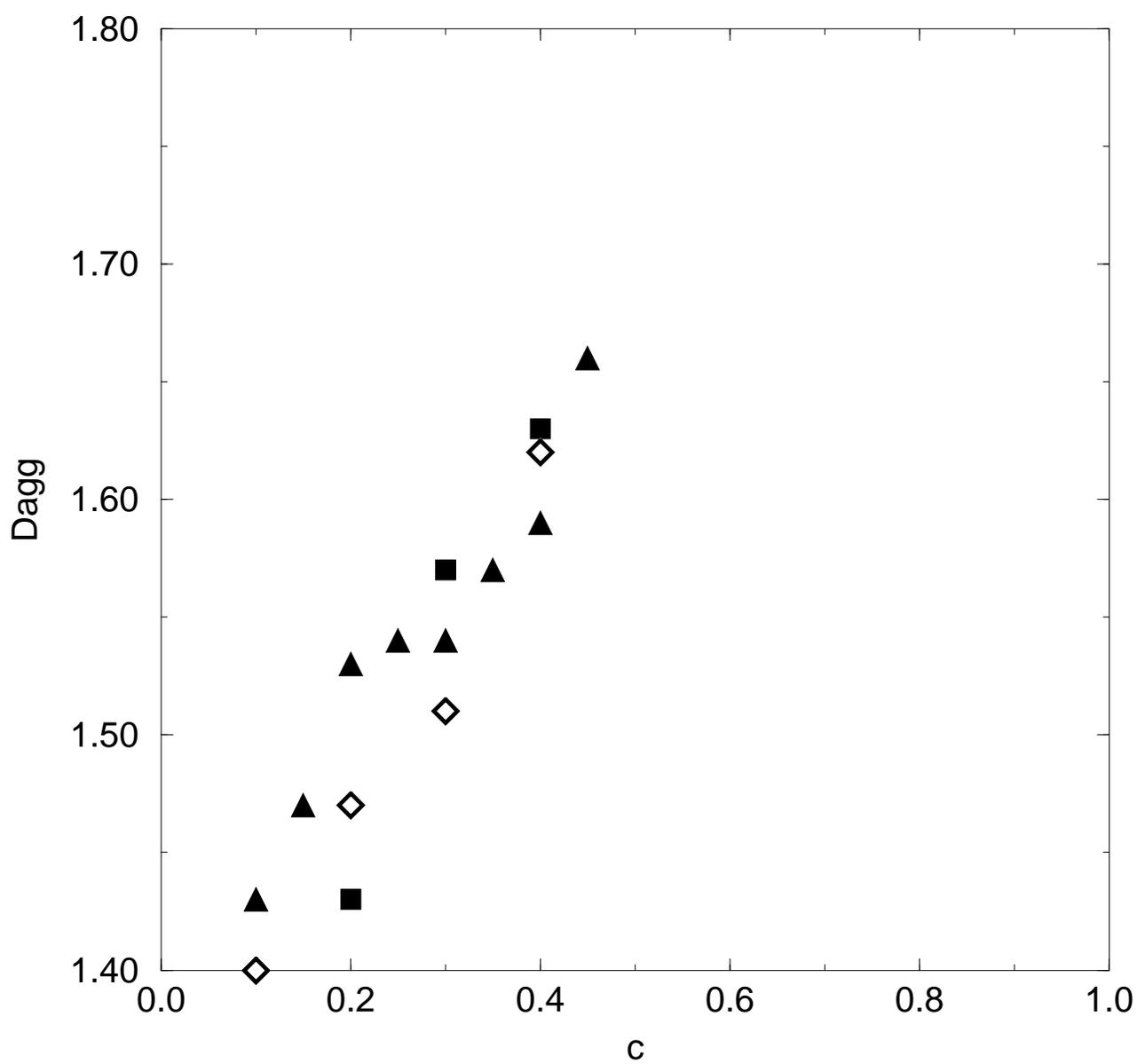
\begin{figure}
\caption{Fractal dimension of the aggregates $D_{agg}$ ($l<L_c$) versus c,
for $L=90$ (black squares), $L=120$ (black triangles) and $L=240$
(open diamonds), resulting from averages over 80, 60 and 40 simulations, respectively.}
\end{figure}

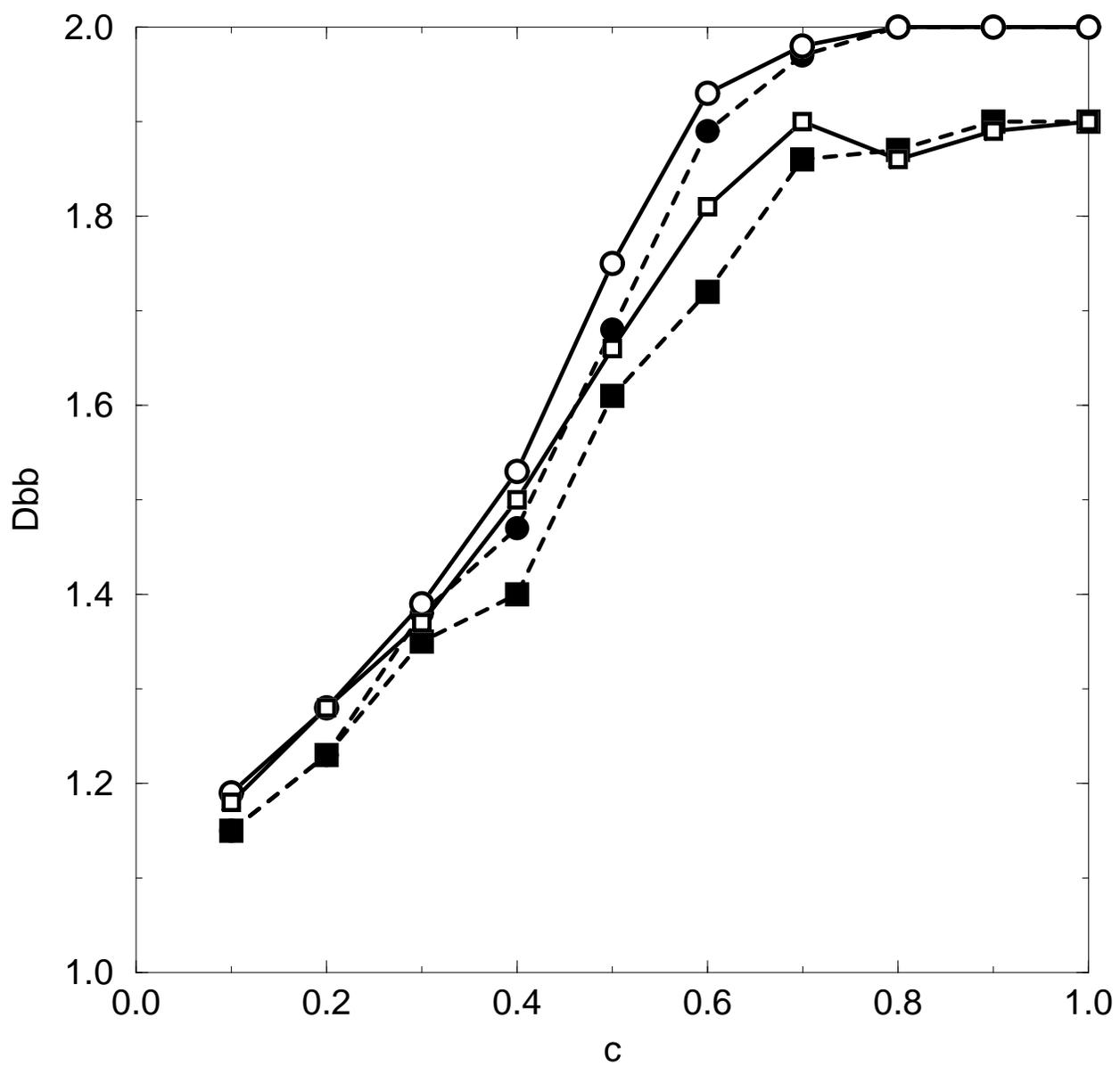
\begin{figure}
\caption{Fractal dimension of the backbone $D_{bb}$ versus c,
for $L=90$ (black symbols) and $L=120$ (open symbols).
Square symbols and circle symbols denote $D_{bb}$ for the gelling cluster
and the final cluster, respectively. These data results from averages
over 80 and 60 simulations, respectively.}
\end{figure}

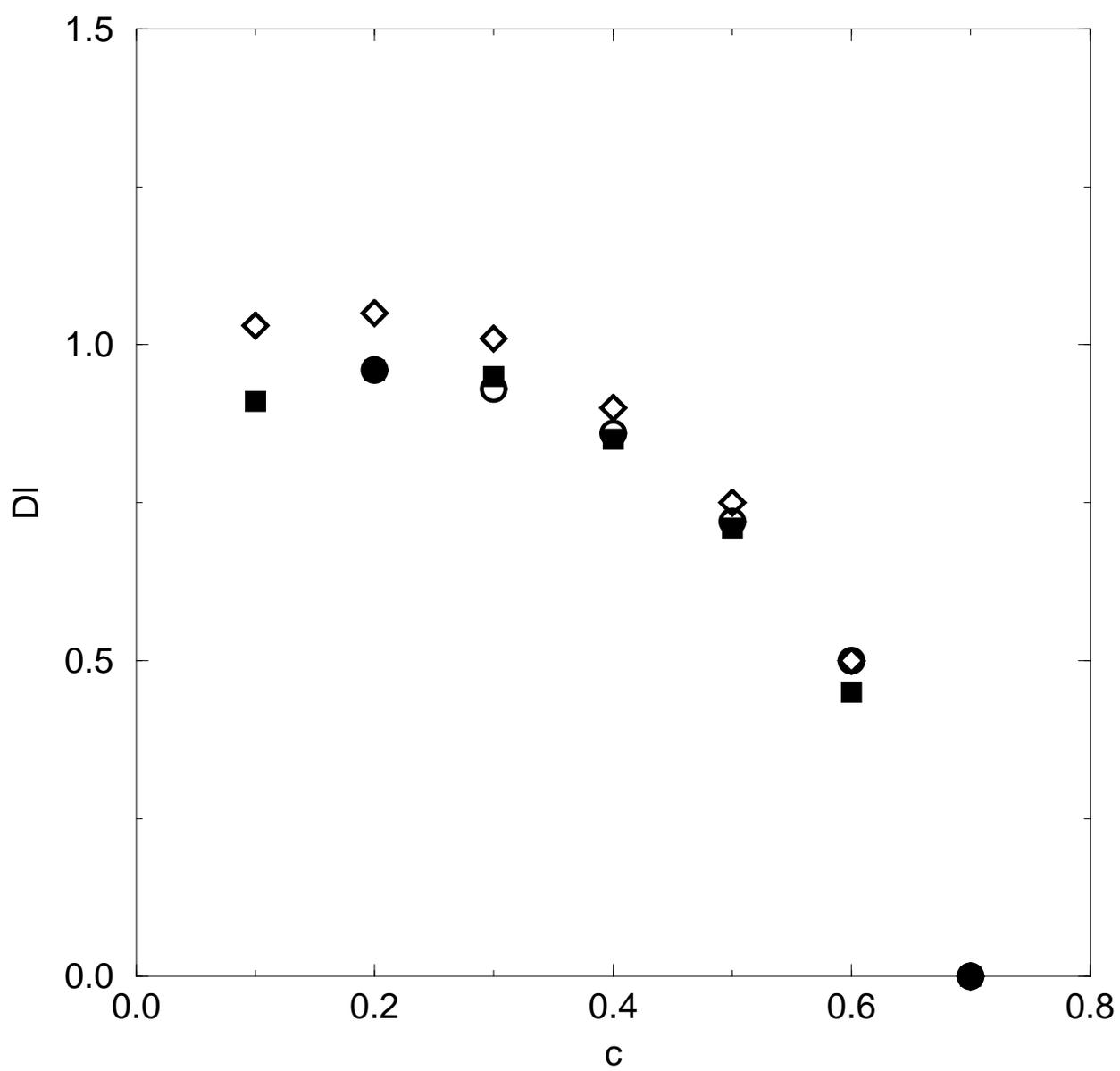
\begin{figure}
\caption{Fractal dimension of the links $D_l$ ($l<L_c$) versus c,
for $L=60$ (open circles), $L=90$ (black squares) and $L=120$
(open diamonds), resulting from averages over 120, 80 and 60 simulations, respectively.}
\end{figure}

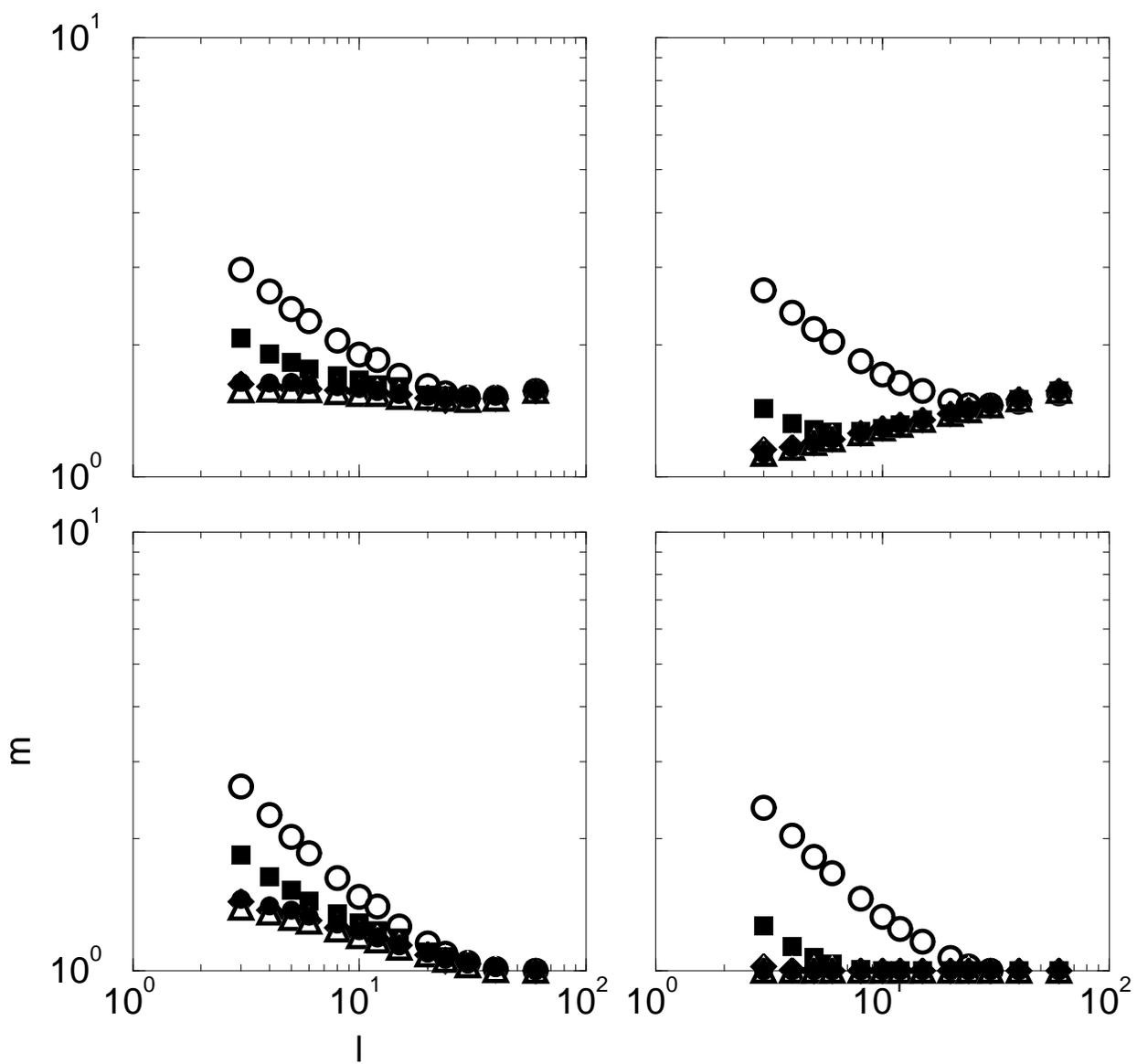
\begin{figure}
\caption{Log-Log plot of $m(={M(l) /  c l^D})$ versus $l$ for $L=120$ and
different concentration values (0.2,0.4,...,1, from top to bottom)
for the RLCA case (with a sticking probability equal to 0.005).
(a) and (c) correspond to G-cluster for $D=D_p$ and $D=d$, respectively.
(b) and (d) correspond to F-cluster for $D=D_p$ and $D=d$, respectively.
These curves result from averages over 20 simulations.}
\end{figure}

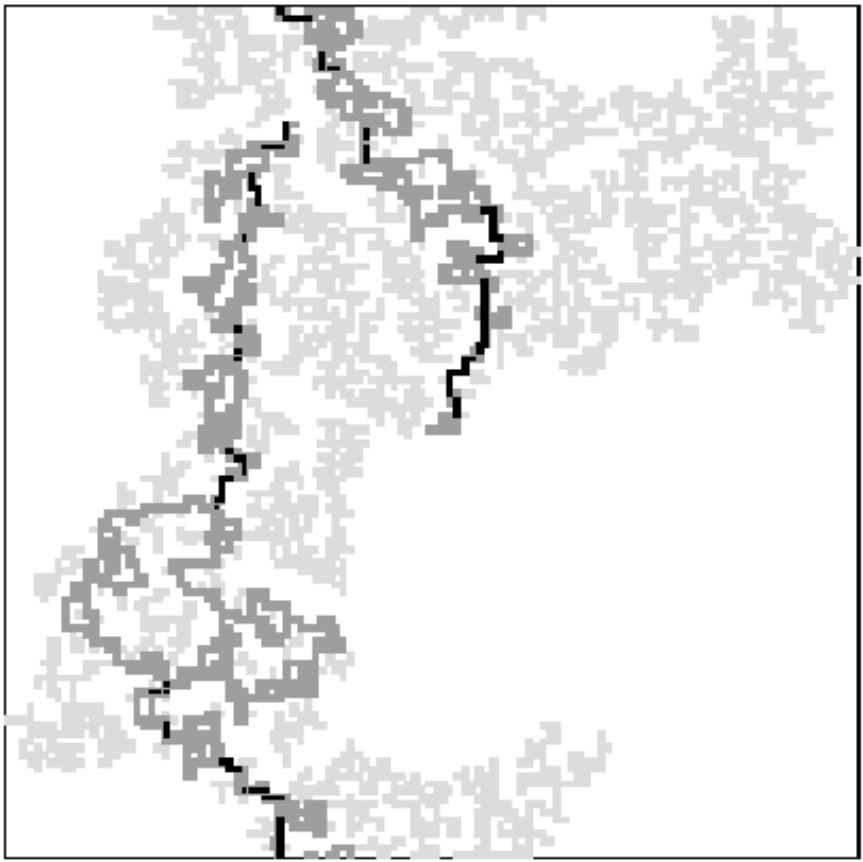
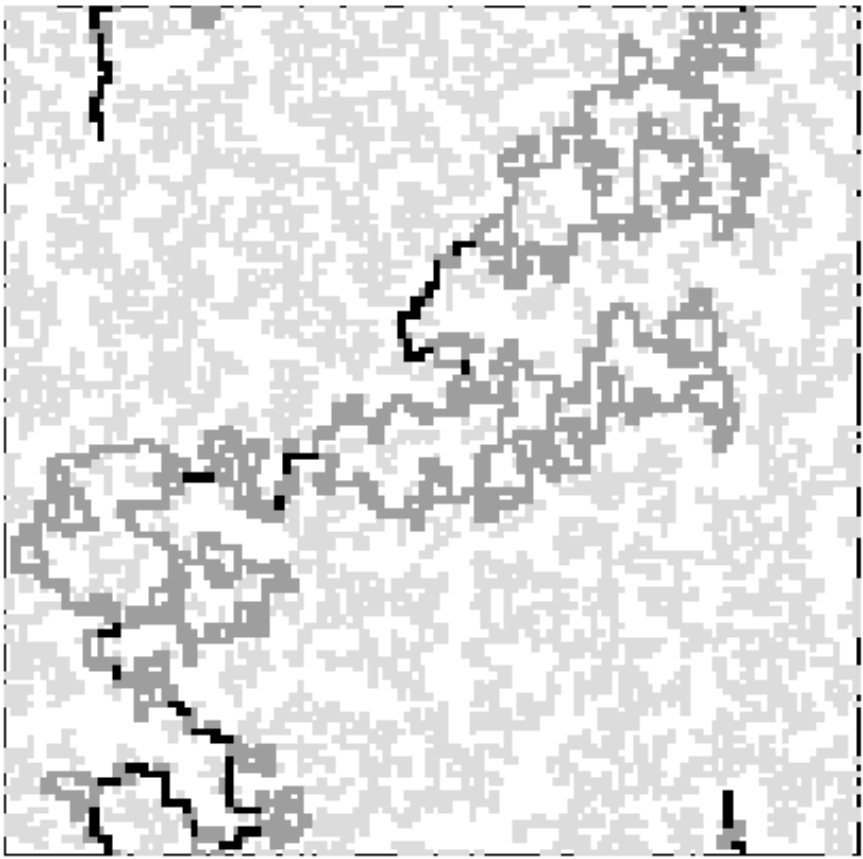
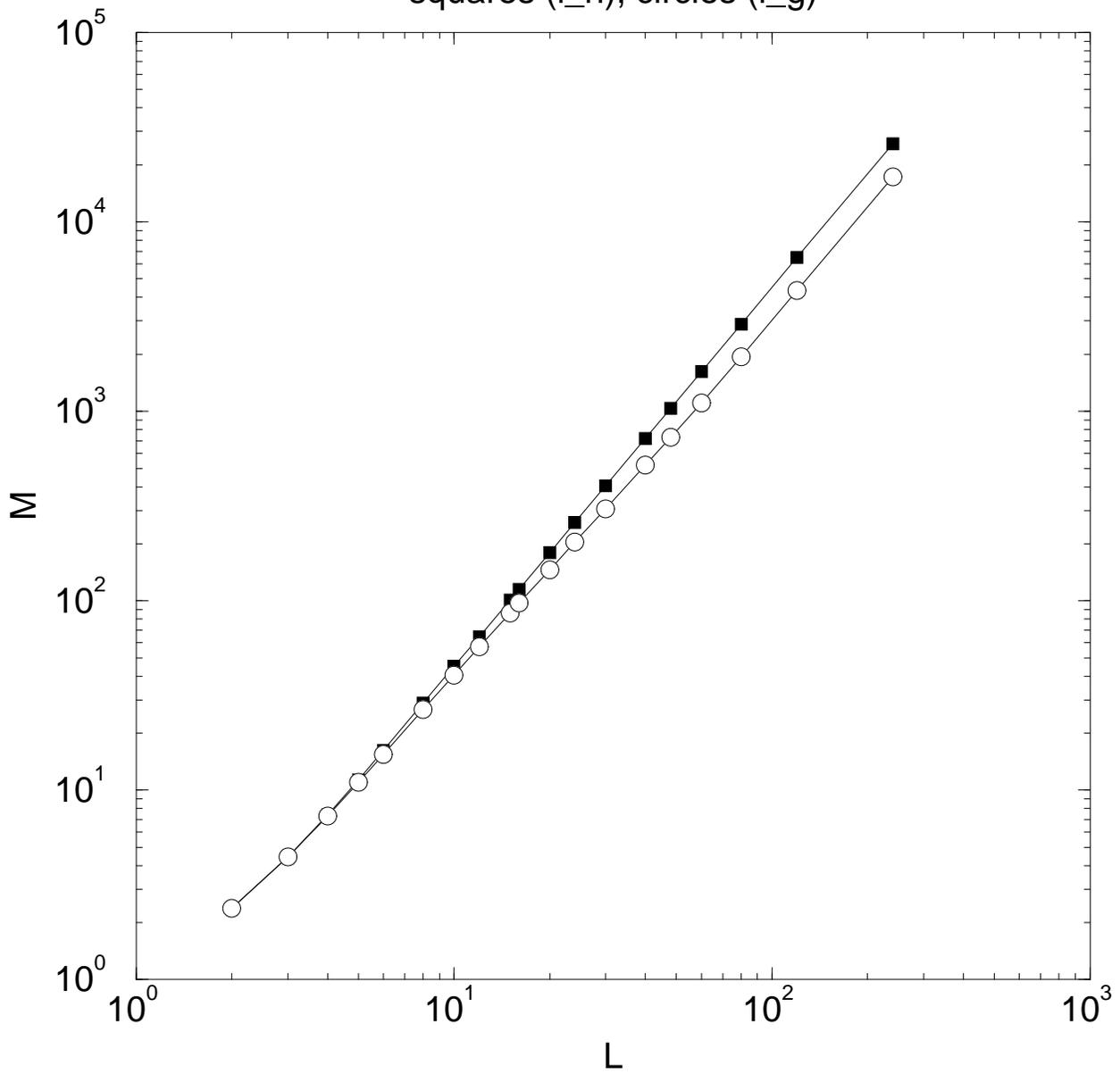
\end{document}